# THE VST TELESCOPE CONTROL SOFTWARE IN THE ESO VLT ENVIRONMENT

P. Schipani, M. Brescia, D. Mancini, L. Marty, G. Spirito
Osservatorio Astronomico di Capodimonte, Napoli, Italy

Abstract

The VST (VLT Survey Telescope) is a 2.6 m Alt-Az telescope to be installed at Mount Paranal in Chile, in the European Southern Observatory (ESO) site. The VST is a wide-field imaging facility planned to supply databases for the ESO Very Large Telescope (VLT) science and carry out stand-alone observations in the UV to I spectral range. This paper will focus mainly on control software aspects, describing the VST software architecture in the context of the whole ESO VLT control concept. The general architecture and the main components of the control software will be described.

## 1  ARCHITECTURE

### 1.1  Maintenance requirements

The VST telescope will be installed at Cerro Paranal in the Atacama desert, Northern Chile, as a powerful survey instrument for the largest ground-based telescope of the world, the ESO VLT. It is a joint project between the Osservatorio Astronomico di Capodimonte (OAC) and ESO. OAC has been committed to design and install the telescope, software included, but after the installation and commissioning VST will be managed by ESO staff. Therefore in order to simplify the future maintenance of the software by ESO staff, the easiest way is to develop the VST software in the most "VLT-compliant" way.

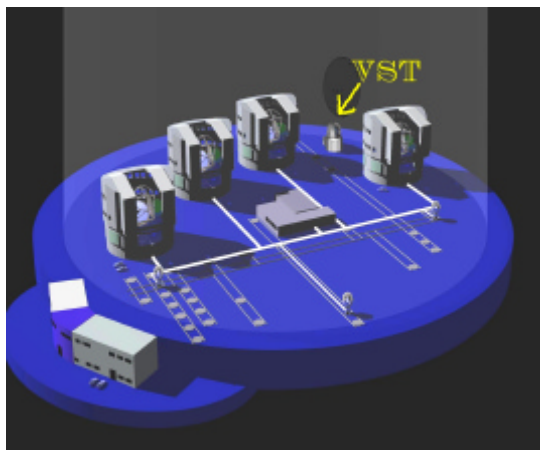

Figure 1: VST in the Paranal observatory (ESO courtesy, http://www.eso.org)

### 1.2  General architecture

As the VLT is a large project involving many staff people for many years, strong standardization has been introduced by ESO in the software development and hardware platforms.

The VST software is based on the same standard architecture: distributed workstations running HP-UX and VME-bus-based Local Control Units (LCU) running the VxWorks real time multitasking operating system, connected together via Local Area Networks.

LCUs interface with field electronics and electro-mechanics, driving encoders, axis controllers, limit switches, etc.; the application software they run is a low level software program, used to control and monitor the hardware devices.

The workstations coordinate LCUs and run the graphical user interfaces (GUI), running high level application software; this kind of software has time execution requirements less urgent than the LCU's one, and is the one the operators usually interface with.

The basic ideas in the architecture are modularity and standardization. Each LCU is devoted to a specific task, so the hardware devices are driven independently by different LCUs, but they use whenever possible the same VME boards and the same software.

The programming languages on HP-UX platform are C++ for the core control modules, and Tcl/Tk for graphical user interface development. The Tcl language is interpreted, so Tcl applications normally do not run as fast as equivalent C programs. However, for a large class of applications, this is not a disadvantage since the processing speed of modern computer systems is more than adequate. The LCU modules are developed in C.

A key role in both workstation and LCU software architecture is played by an On Line Data Base, which on-line allows sharing information between modules, monitoring the status of the system and the execution of commands. The database is built using HP RTAP (Real Time Application Platform), available on HP-UX workstations. RTAP is a modular, Unix based software platform for real-time data acquisition and control of continuous processes. It is a Supervisory Control and Data Acquisition (SCADA) system.

Between the operating system and the application layer there is a software layer developed by ESO for VLT which is reused for VST, the VLT Common Software, providing a lot of utilities, tools and common services to the application programmer.

## 2 TELESCOPE CONTROL SOFTWARE

### 2.1 Introduction

The VST Telescope Control Software (TCS) is the software package responsible for:

- Control and monitoring of the telescope
- Interface to users and other VST software packages
- Interface to star catalogues

### 2.2 Module structure

Each TCS module is provided with a Command Definition Table (CDT) which is normally used for all inter-module communications and is also used by the TCS user interface panels.

In the TCS general module architecture all the modules are implemented through one or more independent processes. If a module is made up of more than one process only one of them is the "control" process which provides the public interface, receives commands from other modules and sends back the corresponding replies. In order to respect these requirements every process is designed in an event-driven way and the implementation is based on the event-handler design. The TCS package is composed by several modules associated to the telescope subsystems to be handled.

Each module is organized in processes and functions dedicated to the LCU and/or WS resident applications. The applications use specific commands to perform all the actions foreseen for the control and management of the associated subsystem.

### 2.3 TCS: VST vs VLT

Due to the maintenance requirements, there is a commitment to the VST team in agreement with ESO, to re-use as much as possible the existing VLT Telescope Control Software.

Since VLT and VST have many similarities in the coordination and control but are very different in many hardware solutions (they have very different size and scientific purposes), this is usually much more possible in the workstation part than in the LCU one, which highly depend on the hardware used in the controlled devices.

The main physical devices to be controlled actually are:

- azimuth
- altitude
- Cassegrain instrument rotator
- hydrostatic bearing support system
- motor cooling system
- adapter
- probe
- focusing device
- pick-up mirror
- atmospheric dispersion corrector
- two-lens corrector
- guiding camera
- Shack-Hartmann device
- M1 support system
- M2 support system

Comparing the VST mechanical and hardware solutions with the VLT ones device by device (and considering there are some new devices in VST not present in VLT) it becomes clear that in the LCU software the percentage of the VLT TCS that can be reused is very low, because the differences are often so large that it is convenient to re-write the related software modules from scratch. The leading concept is that wherever possible the interface (i.e. the Command Definition Table, see 2.2) with the outside world of the new LCU modules is structured in order to be compatible with the existing and reusable (when this is the case) workstation modules. A couple of examples are briefly described in the following.

### 2.4 The case of M1 pad control

One of the VST subsystems whose hardware is completely different from the VLT one is the M1 pad control system. The primary mirror is supported by 84 axial and 24 radial pads. The mirror sag can be modified applying a set of forces to the astatic levers in order to correct $3^{rd}$ and $4^{th}$ order aberrations as in the active optics scheme usual in the new generation of telescopes.

The control system used in VST to implement the pad control is based on a CAN bus network of custom developed microcontroller boards. The introduction of CAN bus is a new feature for the Paranal observatory, so there were no standards to follow either for hardware or software. Consequently the software module to exchange information with the astatic lever controllers, following an established communication protocol, is completely new, even if developed following the same structure and general rules of all the other modules.

## 2.5 The case of Atmospheric Dispersion Corrector control

The atmospheric dispersion correction in VST is obtained actually using two different optical correctors, user selectable depending on the altitude angle: they are named "ADC" and "2-lens corrector". The ADC is composed by two rotating pairs of prisms, while the 2-lens corrector optics positions are fixed.

Here the difference with VLT basically in the usage of two different correctors, in the kind of motion (circular and not linear) of the optics, in the algorithm to calculate the prism positions, and in the kind of the selected motors. Therefore even if at first glance there are many similarities in the two systems, the LCU modules must be completely different, and the workstation module needs some modifications. Again, the old interface between modules (CDT) can be largely saved, and extended with new features where needed.

## 3 REFERENCES

[1] D. Mancini, G. Sedmak, M. Brescia, F. Cortecchia, D. Fierro, V. Fiume Garelli, G. Marra, F. Perrotta, P. Schipani, : 2000, "VST project: technical overview", in "Telescope Structures, Enclosures, Controls, Assembly / Integration / Validation, and Commissioning", eds. Sebring, T., Andersen, T., SPIE, 4004, pp. 79-90

[2] D. Mancini, G. Sedmak, M. Brescia, F. Cortecchia, D. Fierro, V. Fiume Garelli, G. Marra, F. Perrotta, P. Schipani, : 2000, "VST Final Design Review", http://vst.na.astro.it

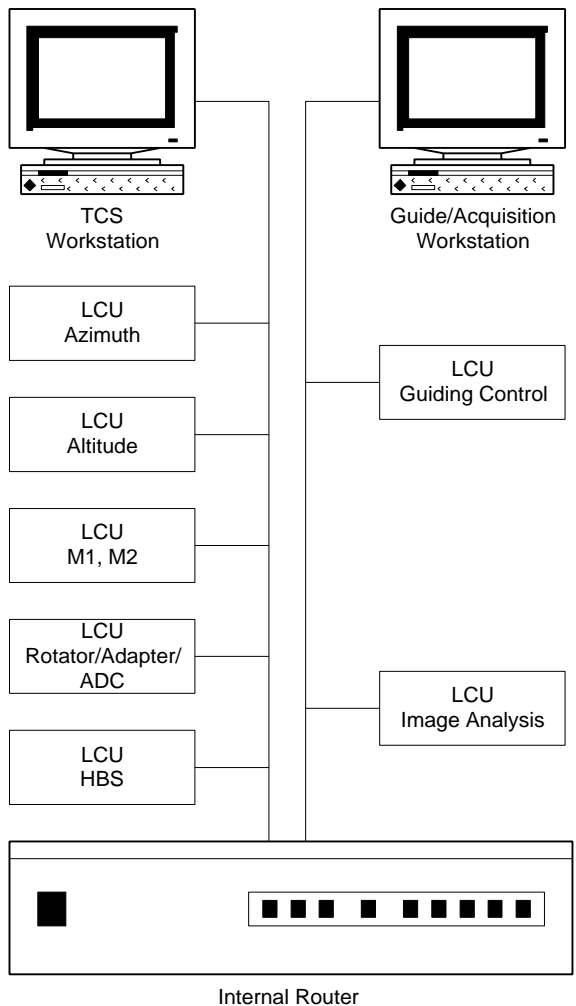

Figure 2: Telescope Control Architecture